\documentclass[twocolumn,showpacs,preprintnumbers,amsmath,amssymb,superscriptaddress]{revtex4}

\usepackage{graphicx}
\usepackage{dcolumn}
\usepackage{bm}

\begin{document}


\title{Temperature square dependence of the low frequency $1/f$ charge noise in the Josephson junction qubits}

\author{O. Astafiev}
\email{astf@frl.cl.nec.co.jp}
\author{Yu. A. Pashkin}
\altaffiliation[On leave from ]{Lebedev Physical Institute, Moscow
117924, Russia}
\author{Y. Nakamura}
\author{T. Yamamoto}
\author{J. S. Tsai}
\affiliation{NEC Fundamental and Environmental Research
Laboratories, Tsukuba, Ibaraki 305-8501, Japan} \affiliation{The
Institute of Physical and Chemical Research (RIKEN), Wako, Saitama
351-0198, Japan}

\date{\today}

\begin{abstract}
To verify the hypothesis about the common origin of the low
frequency $1/f$ noise and the quantum $f$ noise recently measured
in the Josephson charge qubits, we study temperature dependence of
the $1/f$ noise and decay of coherent oscillations. $T^{2}$
dependence of the $1/f$ noise is experimentally demonstrated,
which supports the hypothesis. We also show that dephasing in the
Josephson charge qubits off the electrostatic energy degeneracy
point is consistently explained by the same low frequency $1/f$
noise that is observed in the transport measurements.
\end{abstract}

\maketitle

Due to potential scalability, Josephson quantum bits are good
candidates for building quantum computers \cite{Squbit}. To have a
long decoherence time, the qubits should be well decoupled from
all noise sources, in particular, charge noise from uncontrollable
charge fluctuations. Therefore, the noise and decoherence in the
qubits are now the key issue of the qubit research. Although the
noise has been studied in a number of works \cite{f-noise,
Nakamura-dephasing, Saclay, Schoelkopf, Bertet, Martinis2}, its
nature is still not yet understood.

The low frequency noise in metallic single electron transistors
(SETs) has been studied intensively a while ago \cite{Martinis,
Mooij, Krupenin}. It has been found that the noise is produced by
charge fluctuators and its spectral density is close to $1/f$.
Recently, it has been shown that the low frequency charge noise
gives the main contribution to dephasing of coherent oscillations
in the Josephson charge qubits \cite{Nakamura-dephasing, Saclay}.
The high frequency quantum noise of the environment has been
investigated in Ref. \cite{f-noise} by using a qubit as a quantum
spectrometer. The qubit relaxation is caused by the asymmetric
quantum noise with a non-monotonic spectrum, which tends to have a
linear frequency dependence ($f$-noise) in a wide energy range
(from 2 to 100 GHz$\times h$) like in the case of a simple ohmic
environment. The quantum noise originates from absorption of the
energy of excited qubits by the cold environment and, therefore,
should be nearly temperature independent in the range of qubit
energies higher than $k_{B}T$. Surprisingly, it turned out that
the amplitude of the $f$ quantum noise crosses the low frequency
$1/f$ noise extrapolated to the gigahertz range at a frequency
$\omega_{c}\sim k_{B}T/\hbar$, which implies that both noises may
have a common origin. Furthermore, if one assumes that the
crossover frequency $\omega_{c}$ scales linearly with temperature
then the strength of the $1/f$ charge noise should be proportional
to temperature square ($T^{2}$ dependence) \cite{f-noise}.
Although $T^{2}$ dependence has been observed in the critical
current noise of the Josephson junction \cite{Wellstood}, and
maybe related to charge fluctuations in the junction, the
dependence appears to be unexpected in the charge noise
measurements as it contradicts to the linear temperature
dependence of the $1/f$ noise in glasses \cite{Dutta, Kogan}.
Therefore, the $T^{2}$ dependence in the charge $1/f$ noise has to
be experimentally confirmed. In addition, temperature dependence
gives important information on the density of states of the
fluctuators and will help to verify theoretical models of the
$1/f$ noise intensively studied recently in Refs. \cite{Makhlin,
Lara1, Lara2, Lerner, Sousa}. Some of these works are based on the
prediction of the $T^{2}$ dependence reported in Ref.
\cite{f-noise}. Unfortunately, temperature dependence of the $1/f$
noise in SETs has not been studied in details in earlier works. It
was found that the noise increases with temperature and saturates
in the low temperature range \cite{PTB, Maryland}. In Ref.
\cite{Maryland}, the quadratic temperature dependence was expected
but has not been actually demonstrated.

In this work, we study temperature dependence of the $1/f$ noise
in the Josephson charge qubits by measuring dc transport in the
SET regime. We have found that the noise exhibits
$T^{2}$-dependence at temperatures from 200 mK up to $\sim$ 1 K.
We also study the effect of temperature on dephasing of the qubit
during coherent oscillations. Decay of the coherent oscillations
away from the degeneracy point of electrostatic energy is
consistently explained by dephasing due to the low frequency $1/f$
charge noise. We also briefly discuss phenomenological models of
$T^{2}$ dependences from the experimental point of view.

To study temperature dependence we fabricate qubits with the same
geometry and junction properties as the qubits used in Ref.
\cite{f-noise}. The Al structure is fabricated on top of 400 nm
thick Si$_{3}$N$_{4}$ insulation layer deposited on a gold ground
plane. The total capacitance of the qubit island is about 600 aF
(the charging energy is $E_{C} = e^{2}/2C \approx 30$ GHz$\times h
$) and is mainly formed by its Josephson junction. Instead of the
trap island used in \cite{f-noise}, we fabricate an electrical
lead connected to the qubit through a small highly resistive
tunnel junction with a resistance of 10 $-$ 50 M$\Omega$ (as it
was in our earlier works \cite{Nakamura-dephasing, Saclay}) to
measure current through the qubit in the SET regime.

\begin{figure}
\includegraphics{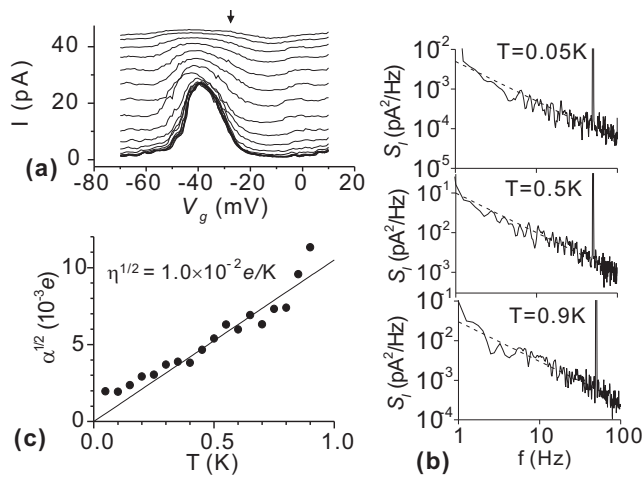}
\caption{\label{fig:1fFig1}(a) A Coulomb peak of the SET measured
at temperatures from 50 to 900 mK with step of 50 mK. (b) Examples
of the current noise spectra $S_I$ at different temperatures
measured on the slope of the SET peak. Dashed lines are the
$1/f$-dependence. (c) Temperature dependence of amplitudes
$\alpha^{1/2}$ of the $1/f$ noise spectra. The solid line is
$\alpha^{1/2} = \eta^{1/2}T$ with $\eta = (1.0 \times
10^{-2}e/$K$)^2$. }
\end{figure}

We use the qubit as an SET and measure the low frequency charge
noise, which causes the SET peak position fluctuations.
Temperature dependence of the noise is measured from the base
temperature of 50 mK up to 900 $-$ 1000 mK. The SET is normally
biased to $V_{b} = 4\Delta/e$ ($\sim$ 1 mV), where Coulomb
oscillations of the quasiparticle current are observed. Figure
\ref{fig:1fFig1}(a) exemplifies the position of the SET Coulomb
peak as a function of the gate voltage at temperatures from 50 mK
up to 900 mK with an increment of 50 mK. The current noise
spectral density is measured at the gate voltage corresponding to
the slope of the SET peak (shown by the arrow), at the maximum (on
the top of the peak) and at the minimum (in the Coulomb blockade).
Normally, the noise spectra in the two latter cases are frequency
independent in the measured frequency range (and usually do not
exceed the noise of the measurement setup). However, the noise
spectra taken on the slope of the peak show nearly $1/f$ frequency
dependence (see examples of the current noise $S_{I}$ at different
temperatures in Fig.~\ref{fig:1fFig1}(b)) saturating at a higher
frequencies (usually above 10 $-$ 100 Hz depending on the device
properties) at the level of the noise of the measurement circuit.
The fact that the measured $1/f$ noise on the slope is
substantially higher than the noises on the top of the peak and in
the blockade regime indicates that the noise comes from
fluctuations of the peak position, which can be translated into
charge fluctuations in the SET.

To obtain the $1/f$ charge noise spectral density
\begin{equation}
S_{q}(\omega) = \frac{\alpha}{\omega} \label{eq:Eq1}
\end{equation}
(defined for frequencies $\omega > 0$) we first take the low
frequency part of the current noise spectral density $S_{I}(f)$
and find the parameter $A$ of the fitting curve $A/f$ as $A =
\langle S_{I}(f)\rangle /\langle 1/f \rangle$
\cite{Noise_derivation}. Next, we transform the current noise into
the charge noise $\alpha = A/(dI/dV_{g})^{2}/(\Delta V_{g}/e)^{2}$
using the transfer function $dI/dV_{g}$ on the slope of the peak
at the measurement point, where $\Delta V_{g}$ is the spacing in
gate voltage between two adjacent peaks (corresponding to the
change of the SET charge by $e$). Dimensionality of $\alpha$ is
$e^{2}$  and a typical value of $\alpha$ is of the order of
(10$^{-3}$ $e$)$^{2}$ at $T \leq$ 200 mK.

Solid dots in Fig.~\ref{fig:1fFig1}(c) represent $\alpha^{1/2}$ as
a function of temperature. $\alpha^{1/2}$ saturates at
temperatures below 200 mK at the level of $2 \times 10^{-3} e$ and
exhibits nearly linear rise at temperatures above 200 mK with
$\alpha^{1/2} \approx \eta^{1/2} T$, where $\eta \approx (1.0
\times 10^{-2} e/K)^{2}$ (the solid line in
Fig.~\ref{fig:1fFig1}(c)). $T^{2}$ dependence of $\alpha$ is
observed in many samples, though sometimes the noise is not
exactly $1/f$, having a bump from the Lorentzian spectrum of a
strongly coupled low frequency fluctuator. In such cases, switches
from the single two-level fluctuator are seen in time traces of
the current \cite{Martinis}.

\begin{figure}
\includegraphics{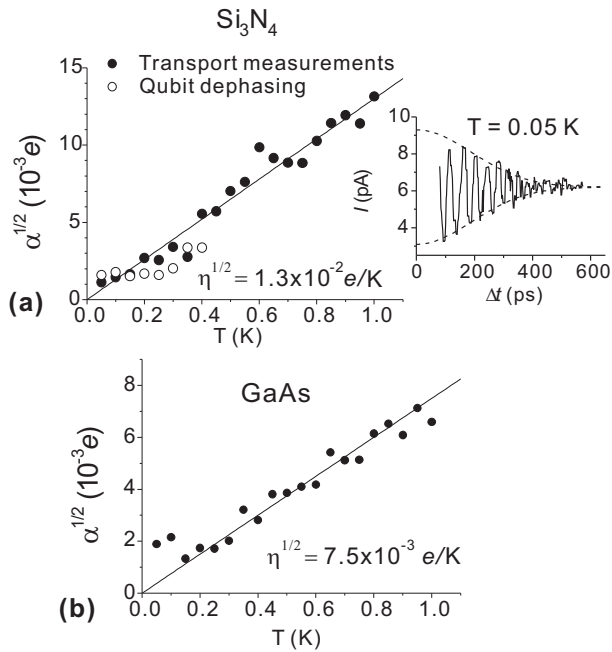}
\caption{\label{fig:1fFig2} (a) Solid dots show temperature
dependence of $\alpha^{1/2}$ with a fixed bias current (the bias
voltage is adjusted to keep the current constant). Open dots show
$\alpha^{1/2}$ derived from the measurement of qubit dephasing
during coherent oscillations. The coherent oscillations (solid
line) as well as the envelope $\exp(-t^2/2T_2^{*2})$ with
$T_{2}^{*}$ = 180 ps (dashed line) are in the inset. (b) Solid
dots show temperature dependence of $\alpha^{1/2}$ for the SET on
GaAs substrate.}
\end{figure}

Note that at a fixed bias voltage the average current through the
SET increases with temperature (see Fig.~\ref{fig:1fFig1}(a)).
However, it has almost no effect on the noise as we confirmed from
the measurement of the current noise dependence. Nevertheless, to
avoid possible contribution from the current dependent noise we
adjust the bias voltage in the next measurements so that the
average current is kept nearly constant at the measurement points
for different temperatures. Fig.~\ref{fig:1fFig2}(a) shows the
temperature dependences of $\alpha^{1/2}$ for a different sample
with a similar geometry taken in the frequency range from 0.1 Hz
to 10 Hz with a bias current adjusted to about $I $ = $12 \pm 2$
pA. The straight line in the plot is $\alpha^{1/2} = \eta^{1/2}
T$, which corresponds to $T^{2}$-dependence of $\alpha$ with $\eta
\approx (1.3 \times10^{-2} e/$K$)^{2}$.

We study decay of coherent oscillations away from the
electrostatic energy degeneracy point at different temperatures by
measuring pulse induced current \cite{Squbit, Nakamura-dephasing,
f-noise}. The Hamiltonian of our qubit written in the charge basis
$|0\rangle$ and $|1\rangle$ (with and without the Cooper pair in
the island) is $H = \frac{\Delta E}{2}(\sigma_z \cos \theta
+\sigma_x \sin \theta)$, where $\theta = \arctan (E_J/\Delta U)$
and $\Delta E = \sqrt{\Delta U^{2} + E_J^{2}}$ are determined by
the electrostatic energy difference $\Delta U$ between the two
charge states of the qubit and the Josephson energy $E_J$. The
electrostatic energy $\Delta U$ is controlled by the qubit gate
voltage. Adjusting a dc gate voltage to the point far away from
the degeneracy ($\Delta U \gg E_J$, $\theta \approx 0$), where the
ground state is nearly $|0\rangle$ we apply a rectangular voltage
pulse of length $t$ bringing the qubit in the vicinity of the
degeneracy point ($\Delta U$ is of the order or smaller than $E_J$
and $\theta$ $\sim \pi/2$), that is the Hamiltonian changes
non-adiabatically to $H_1$ for time $t$. The coherent evolution
can be presented as $\exp(- \frac{i}{\hbar}\int_{0}^{t}H_1
dt)|0\rangle$. Applying a sequence of identical pulses we detect a
pulse induced current, which is proportional to the probability to
find out the qubit in the state $|1\rangle$ after the pulse
manipulation.

The typical current oscillation as a function of $t$ away from the
degeneracy point ($\theta \neq \pi/2$) is exemplified in the inset
of Fig.~\ref{fig:1fFig2}(a). If dephasing is induced by the
Gaussian noise, the oscillations decay as $\exp
(-t^{2}/2T_{2}^{*2})$ with
\begin{equation}
\frac{1}{{T_2^{*2}}}\approx\frac{\cos^2\theta}{\hbar^2}\biggl(\frac{4E_C}{e}\biggr)^2
\int_{\omega_0}^{\infty}S_q({\omega})\biggl( \frac{2\sin(\omega
t/2)}{\omega t}\biggr)^2d\omega, \label{eq:Eq2}
\end{equation}
where $\omega_{0}\approx 1/\tau$ is the low frequency integration
limit defined by the measurement time constant $\tau$. In the case
of the $1/f$ Gaussian noise of Eq.~(\ref{eq:Eq1})
\begin{equation}
T_{2}^{*}\approx\frac{e \hbar}{4E_{C}\sqrt{\alpha
\ln(\omega_{1}\tau)}\cos \theta},
 \label{eq:Eq3}
\end{equation}
where $\omega_{1} \leq \pi/T_{2}^{*}$ is the effective high
frequency limit.

Qubit dephasing due to the non-Gaussian noise is treated in Refs.
\cite{Galperin, Falci}. For instance, in the case of a strongly
coupled fluctuator, the decay is slower than Gaussian. However,
importantly, Eq.~(\ref{eq:Eq2}) used for fitting the initial part
of the oscillations still gives a reasonably good agreement for
estimation of the amplitude of the $1/f$ noise. We have analyzed
time traces of the noise and found that the noise is often close
to the Gaussian, but sometimes it is clearly not, for example, in
the presence of a strongly coupled fluctuator .

The solid line in the inset of Fig.~\ref{fig:1fFig2}(a) shows
decay of coherent oscillations measured at $T = 50$ mK and the
dashed envelope exemplifies a Gaussian with $T_{2}^{*}$ = 180 ps.
We derive $\alpha^{1/2}$ from Eq.~(\ref{eq:Eq3}) and plot it  in
Fig.~\ref{fig:1fFig2}(a) by open dots as a function of
temperature. The low frequency integration limit and the high
frequency cutoff are taken to be $\omega_{0} \approx 2\pi \times
25$ Hz and $\omega_{1} \approx 2\pi \times 5$ GHz  for our
measurement time constant $\tau$ = 0.02 s and typical dephasing
time $T^{*}_{2} \approx 100$ ps \cite{cutoff2}.

The saturation of the $1/f$ noise at low temperatures has also
been observed in earlier works \cite{PTB, Maryland}. Although its
origin is not clear, we can suggest the following possible
mechanisms: (1) heating of an electron system, (2) freezing out
fluctuators, so that the effective number of active fluctuators
decreases down to a few per decade (in this case the $1/f$ noise
saturates to the level of a single fluctuator amplitude).

To collect more information about the $1/f$ noise we perform an
additional experiment studying the noise in the SET fabricated on
a different substrate. Although the results are not conclusive and
require more systematic study we think that it is instructive to
present these data here. Fig.~\ref{fig:1fFig2}(b) demonstrates the
$1/f$ noise temperature dependence for an SET fabricated on
single-crystal GaAs. The GaAs substrate has been chosen to reduce
the number of defects (as a possible origin of fluctuators)
typical for amorphous materials like CVD grown Si$_{3}$N$_{4}$ or
thermal oxide on top of bare silicon. Again, clear
$T^{2}$-dependence is observed above 200 mK, and $\eta \approx
(0.75\times 10^{-2}e/$K$)^{2}$ is lower but close to what was
measured in the case of Si$_{3}$N$_{4}$.

Below we discuss a phenomenological model explaining the
properties of the noise. Our qubit is coupled to charge dipoles
$ed$ in the insulator, which, in turn, induce a charge $\delta q$
in the qubit island (the fluctuators at distances smaller than the
characteristic size of the island $R$ produce $\delta q \sim
ed/R$, and the matrix element of the Cooper pair transition is at
most $4E_{C}\delta q$). Based on the phenomenology from Ref.
\cite{f-noise}, it has been pointed out in Refs. \cite{Makhlin,
Lara2} that the $T^{2}$ dependence may originate from two-level
fluctators (characterized by bias energy $\epsilon$ and tunneling
energy $\Delta$) and linearly distributed in $\epsilon$ with an
energy independent amplitude $\langle \delta q \rangle$. Summation
of the Lorentzian spectra over many fluctuators with energies
$\epsilon$ below $k_{B}T$ gives $(k_{B}T)^{2}$ term in the noise
spectrum. Such linear energy distribution, for example, appears in
the models treating charge fluctuations between the
superconducting island and an insulator \cite{Lara1, Lara2}. The
switching rate of the thermally activated fluctuators ($\epsilon <
kT$) can be presented as $\gamma \sim
\gamma_{0}(\Delta/\epsilon)^{2}$, where $\gamma_{0}$ depends on
the coupling of the fluctuators to the external thermal bath, and
$\Delta$ is the tunneling energy of electrons in the fluctuators
\cite{cutoff1}. The slow fluctuators, contributing to the
low-frequency $1/f$ noise should have a strong suppression factor
$(\Delta/\epsilon)^{2} \ll 1$. On the other hand, to efficiently
absorb qubit energy, the two-level systems should have $\Delta$ of
order or larger $\epsilon$. Note that the two-level systems
producing the $1/f$ noise and absorbing the qubit energy are
characterized by very different values of $\Delta$. A commonly
used assumption that two-level systems are distributed according
to $P(\Delta) \propto 1/\Delta$ gives rise to the crossover
frequency of $\omega_{c} \approx k_{B}T/\hbar$ \cite{Makhlin}.

To show relationship of the model with experiments we provide some
numbers from experiments. The $1/f$ noise in this work is studied
in the temperature range from 0.05 to 1 K, which means that only
fluctuators with the activation energies lower than $k_{B}T$ (from
1 to 20 GHz$\times h$) produce the noise. Note that this thermal
energy range overlaps with the qubit energy 2 $-$ 100 GHz$\times
h$ for which the $f$ quantum noise has been studied in Ref.
\cite{f-noise}. On the other hand, the $1/f$ noise is measured in
the frequency range 0.1 $-$ 100 Hz, which gives typical values of
$\gamma$ for the fluctuators contributing to the measured noise.
The high frequency cutoff of the $1/f$ noise (which may give rough
estimate of $\gamma_{0}$) is not known for our qubits. For rough
estimations, we take $\gamma_{0}$ = 1 MHz from Ref. \cite{Saclay}
and find that only fluctuators with $(\Delta/\epsilon)^2 \approx
\gamma/\gamma_{0} \sim 10^{-7} - 10^{-4}$ contribute to the
measured noise. Note that to have necessary relationship between
the fluctuators and the two-level system absorbing the qubit
energy, distribution $P(\Delta) \propto 1/\Delta$ should hold in a
very wide range of $\Delta$ from 10$^{6}$ Hz$\times h$ or less up
to 10$^{11}$ Hz$\times h$.

Apart from the phenomenology, the most important question now is
what the detailed mechanism of the $1/f$ and $f$ noises is. A few
microscopic models have been proposed in \cite{Lara1, Lara2,
Sousa}. Test experiments have to be done to verify the models. For
instance, behavior of the noise as a function of magnetic field
and superconducting-normal state transition would be important for
the theories involving superconductivity.

In conclusion, we have observed $T^2$ temperature dependence of
the low frequency $1/f$ noise, which supports the idea that the
$1/f$ and $f$ noises have a common origin. Typically, the noise
spectral density is $(10^{-2} e)^2(T/$K$)^2/\omega$ at $T
> 200$ mK and saturates to the level of the order of $(10^{-3}
e)^2/\omega$ at $T < 200$ mK. We demonstrated that free induction
decay is consistently explained by dephasing on the low frequency
$1/f$ noise. $T^2$ dependence of similar amplitude is observed for
two different substrate materials amorphous Si$_3$N$_4$ and
single-crystal GaAs.

We thank B. Altshuler, Y. Makhlin, A. Shnirman, G. Sch{\"{o}}n, L.
Faoro and L. Ioffe for valuable discussions. This work has been
partially supported by the Japan Science and Technology
Corporation.


\begin{thebibliography}{25}
\bibitem[{Noi()}]{Noise_derivation}
\bibinfo{note}{To account for the contribution from equal frequency bands
  $\Delta f$ with weight $\Delta f/f$ (log-uniform in frequency) we take mean
  value of $S(f) f$ with weight function $1/f$ (that is $\langle S(f)\rangle$)
  and normalize by the mean value of $1/f$.}
\bibitem[{cut({\natexlab{a}})}]{cutoff2}
\bibinfo{note}{The echo experiment in Ref. \cite{Saclay} shows that the $1/f$
  noise has a high frequency cutoff, which is estimated to be 1 MHz. In the
  case of $\omega_{1} \approx 2\pi\times 1$ MHz, the derived $\alpha^{1/2}$
  will be only 30\% higher.}
\bibitem[{cut({\natexlab{b}})}]{cutoff1}
\bibinfo{note}{$\gamma_{0}$ can depend on temperature and defines the high
  frequency cutoff of the $1/f$ noise, which can be estimated from measuring
  spin-echo like signals \cite{Saclay}.}
\end{thebibliography}

\end{document}